# PRICING AND HEDGING DERIVATIVE SECURITIES WITH UNKNOWN LOCAL VOLATILITIES


Kerry Fendick

fendick@att.net


## Abstract


A common assumption in financial engineering is that the market price for any derivative coincides with

an objectively defined risk-neutral price – a plausible assumption only if traders collectively possess

objective knowledge about the price dynamics of the underlying security over short time scales. Here we

assume that traders have an objective knowledge about the underlying security's price trajectories only

for large time scales. We show that avoidance of arbitrage that is still feasible uniquely determines the

prices of options with large expiration times, and we derive limit theorems useful for estimation of

model parameters and present-value analysis of derivative portfolios.


## Key Words

Risk neutrality; local volatility models; time-inhomogeneous diffusion processes; wide-sense Markov

property; efficient market hypothesis; subjective probabilities





# 1   Introduction

A common assumption in financial engineering is that market prices of derivatives coincide with objectively defined *risk-neutral* prices. According to the first fundamental theorem of asset pricing, risk neutral prices guarantee the absence of all forms of arbitrage, and the absence of arbitrage implies the existence of risk-neutral prices for the given market. When risk-neutral prices are unique (in so called *complete* markets), they can be calculated given objective knowledge about the price dynamics of the underlying securities. Then, according to the second fundamental theorem of asset pricing, any divergence from those risk neutral prices, however small, would result in arbitrage opportunities via dynamic portfolio replication, by which a trader can exactly replicate the payout of a derivative. For background on the fundamental theorems of asset pricing, see Harrison and Kreps [1], Dalang, Morton, and Willinger [2], Delbaen and Schachermayer [3], Rogers [4] [5], Chapter 5 of Shreve [6], and Chapter 6 of Duffie [7].

Although the fundamental theorems of asset pricing hold generally, their connection with market prices is tenuous. To see why, it is instructive to consider a model originally used by Merton [8] in developing the Black-Scholes-Merton formula for option pricing. Merton assumed that the instantaneous return on the underlying security was described by a one-dimensional stochastic differential equation (SDE) in which the instantaneous expected return was a general stochastic process but the instantaneous variance of returns (the square of the local volatility) was a pre-determined function of time.  He further assumed that all traders had a common knowledge of that function, but were free to disagree about the instantaneous expected returns. Footnote 45 of Merton [8] explains,





*It is quite reasonable to expect that traders may have quite different estimates for current (and future) expected returns due to different levels of information, techniques of analysis, etc. However, most analysts calculate estimates of variances and covariances in the same way: namely, by using previous price data. Since all have access to the same price history, it is also reasonable to assume that their variance-covariance estimates may be the same.*

Under those assumptions, Merton showed that portfolio replication was possible. Risk-neutral prices – expressed in closed form through the Black-Scholes-Merton formula - were thereby uniquely defined in a framework in which individual traders based their decisions on their own subjective models of price dynamics of the underlying security (potentially differing in the expected rate of return), but constrained by certain common knowledge about the objective price dynamics (the function describing the instantaneous variance).

In the proof of the Black-Scholes-Merton formula from Merton [8], it was not sufficient for traders to agree on *some* function describing the instantaneous variance: they needed to agree on *the* function objectively describing the instantaneous variance. The above quote from footnote 45 therefore assumed implicitly that returns exhibited some form of stationarity enabling traders to predict future variances and covariances accurately given past variances and covariances.  Empirical studies have not always supported that assumption. As examples, empirical studies of returns from the S&P and Dow Jones composite portfolios by Pagan and Schwert [9] and Phillips and Loretan [10] and from the Euro-dollar exchange by Bassler, McCauley, and Gunaratne [11] found clear evidence of non-stationarity. As Section E.1 of Fama [12] and pages 2-3 of Gatheral [13] have noted, fat-tailed empirical distributions of log returns, commonly observed in empirical studies, also can be symptomatic of non-stationarity.

If an SDE of the form assumed by Merton [8]  were in fact an exact description of the objective price dynamics for the underlying security, but the function describing the future instantaneous variance were





fundamentally unpredictable because of nonstationarity, then unique risk neutral prices would still exist in theory, but traders (and collectively, the "market") would lack the ability to compute them. Trader's might then lack even the knowledge that those objective dynamics were described by an SDE of that form. As Section 5.3.2 on pages 222-223 of Schreve [6] shows, such objective knowledge about the structure of the price dynamics of the underlying security at infinitesimal time scales is required for dynamic portfolio replication. Knowledge, for example, that the objective price dynamics follow an Ito process -- a generalization of a one-dimensional SDE – is insufficient.  Derman and Taleb [14] have argued that dynamic portfolio replication is not commonly used in practice in large part because of uncertainty about the objective dynamics over short time scales.

The above thought experiment shows that the fundamental theorems of asset pricing do not provide a basis to expect that market prices will agree with objective risk-neutral prices without also assuming knowledge by traders of certain aspects of the price dynamics of the underlying security over arbitrarily short time scales. Without such knowledge, arbitrage might still be impossible, but then because of the *inability* of traders to detect and exploit any differences from the objective risk-neutral prices. Given that complete markets (as in our thought experiment) need not converge to objective risk neutral prices, there is no reason to expect that incomplete markets need do so. *The absence of arbitrage constrains and defines market pricing only when arbitrage is otherwise possible given the knowledge possessed by traders.*

In this paper, we study the pricing of derivatives and hedging of volatility risks when the objective dynamics of the underlying security is in large part unknown or even unknowable. We study these problems from the perspective of an individual trader who uses his or her own subjective probability model for the price evolution of the underlying security.  Our prototypical trader is one with the needs





to hedge positions and to price any conceivable type of derivative, including derivatives not offered on exchanges. The derivatives desk of a sell-side bank approximates such a trader.

Without assuming any knowledge by the trader of the objective price dynamics of the underlying security, we first show in great generality that a necessary condition for preventing certain forms of arbitrage by counterparties is risk-neutral pricing of derivatives relative to whatever *subjective* model is used by that trader in assessing the present value of the derivatives' payouts. We then derive properties of derivative pricing further assuming that different traders possess an objective common knowledge only about quadratic variations of the underlying security's price over large intervals. Our results are consistent with scenarios in which security prices lack stationary structure and in which instantaneous variances themselves vary unpredictable with no a priori bounds over short time scales.

We show that if each trader's subjective model of price dynamics of the underlying security can be expressed as an SDE, then the prices at which those different traders are willing to buy or sell a given option must converge -- as the option's expiration time becomes large -- to a price that depends only on the limited common knowledge about properties of price trajectories of the underlying security over large time scales. Otherwise, a trader's counterparties can exploit that common knowledge for arbitrage, even when no one's subjective model for the underlying security's price has any relationship to objective reality over short time scales.

Under the additional assumption that a trader's subjective model of the underlying price process has the property that incremental log returns are wide-sense Markov, we also show that the finite-dimensional distributions of log returns must similarly converge to multivariate log-normal distributions as time scales become large. This central limit theorem leads to natural heuristics for agreeing on the present value of derivatives having long expirations without assuming objective knowledge (or even a consensus) about market behavior over short time scales.





The problem of *uncertain local volatilities* in derivative pricing and hedging was previously addressed in Avellaneda, Levy, and Paras [15] and in several subsequent papers that their work inspired; see Martini and Jacquier [16] for a survey. Their results apply for options with arbitrary expiration times but assume that local volatilities are uncertain within non-trivial bounds. Our results apply only in the limit as expiration times become large, but address the problem of unknown local volatilities for which no such bounds are available.

## 1.1 Price Trajectories of the Underlying Security

For the purposes of this paper, *derivatives* are defined as bets on the future evolution of a security price with payouts that are each determined by that evolution up to the time of the payout's occurrence. The security itself is then called the derivate's *underlying*. We restrict consideration in this paper to derivates with a single common underlying. We will let $S(t) > 0$ denote the price of the underlying at time $t \geq 0$, and let

$$X^{(0)}(t) \equiv \log \frac{S(t)}{S(0)} \tag{1.1.1}$$

denote the log-return at time $t \geq 0$ relative to the security price at time 0. For $t \geq 0$ and $T \geq 0$, the quadratic variation of the sample path of $X^{(0)}$ over the interval $[t, t + T]$ is defined as

$$\left[X^{(0)}\right]_{t,t+T} \equiv \lim_{\|\Pi\| \to 0} \sum_{k=0}^{n-1} \left(X^{(0)}(t_{k+1}) - X^{(0)}(t_k)\right)^2 \tag{1.1.2}$$

where $\Pi = \{t_0, t_1, \dots, t_n\}$ for $t = t_0 < t_1 < \cdots < t_n = t + T$ is a partition, and $\|\Pi\| = max_{k=0,\dots,n-1}(t_{k+1} - t_k)$.

We will assume that there exist positive constants $\alpha$, $\gamma$, $\theta$, and $T_0$ not depending on $t$ or on the particular sequence of partitions in (1.1.2) such that





$$\left| \frac{1}{T} \left[ X^{(0)} \right]_{t,t+T} - \alpha \right| \le \frac{\theta}{T^\gamma} \; for \; T \ge T_0. \tag{1.1.3}$$

Following mathematical convention, we will say that $f(T) = O(g(T))$ on some interval if there exists a constant $\theta > 0$ such that

$$|f(T)| \le \theta |g(T)| \; for \; T \; in \; that \; interval.$$

Then, (1.1.3) implies that

$$\frac{1}{T} \left[ X^{(0)} \right]_{t,t+T} = \alpha \left( 1 + O \left( \frac{1}{T} \right)^\gamma \right) for \; T \ge T_0, \tag{1.1.4}$$

and (1.1.4) implies the existence of a bound of the form (1.1.3). The definition in (1.1.1) is motivated by the implicit assumption that security prices scale exponentially. The quadratic variation as defined in (1.1.2) is otherwise model-independent. The property (1.1.3) makes no assumptions about continuity of log returns, but does implicitly assume that sample paths have infinite variation (an example being a sample path with jumps). Under the additional assumption that returns evolve as a stochastic process, we would interpret (1.1.3) as holding with probability one relative to some underlying probability space. In that context, (1.1.4) would be called a *strong approximation*; see for example the discussion following Theorem 5 of Glynn [17] . But (1.1.3) does not require that log returns are stochastic, and we will assume throughout that (1.1.3) holds for all trajectories.

## 1.2 Overview

This paper addresses two broad questions about the implications of assuming (1.1.3) as the only objective knowledge available to traders about the underlying price process. Section 2 describes how derivative pricing by an individual trader is constrained to prevent forms of arbitrage that are still possible. Section 3 then prescribes how a trader may compute the present value of portfolios containing derivatives with large expirations. Hence, Section 2 is primarily concerned with pricing, and Section 3 with tools for hedging.





In Section 2, we take the perspective of a trader making pricing decisions about derivatives at a snapshot in time given the available history of the underlying's prices. We assume that the trader's pricing decisions are based on a subjective model of how the underlying's prices will evolve from that point forward. We make no assumptions there about how that model was derived and impose no requirement that it is consistent with any model used by the trader to price options at any earlier or later time.

In Section 2.1, we define the trader's subjective model for the underlying's price evolution through a functional describing the trader's expectations about the present values of future payouts from its derivatives. We show how a trader's pricing of derivatives to prevent certain specific forms of arbitrage requires that those prices are risk-neutral relative to the probabilities defining those subjective expectations.

We obtain the results of Section 2.1 in great generality by extending an axiomatic framework from Theorem 1 of Rogers [18]; see also Rogers [5] for additional discussion of the same result. We begin by considering assumptions that are close analogs to those of Rogers and conclude in Proposition 1, as Theorem 1 from Rogers [18] did, that the pricing functional for derivatives is related to the expectation functional through a Radon-Nikodym density process that is strictly positive.

Our treatment of that material removes a restriction from Rogers [18] that a derivative's payout is bounded. This generalization is needed in Section 2.2, where we apply the results of Section 2.1 for an SDE model that includes no explicit upper bound for the price of the underlying. We obtain the generalization by invoking a version of the Riesz Representation Theorem. The proof of Proposition 1 otherwise borrows heavily from the proof of Theorem 1 from Rogers [18].

Section 2.1 then adds an assumption not present in Rogers [5] [18] under which we conclude in Proposition 2 that the pricing functional is indeed risk-neutral relative to the probabilities defining the





trader's expectation functional. We describe there a connection between that additional assumption and the requirement for put-call parity. Derman and Taleb [14] previously showed that a form of risk neutrality was required for consistency with put-call parity under particular log-normal assumptions about the pricing measure.

Our treatment of the material in Section 2.1 differs from Rogers [18] in two other respects. First, we distinguish between relationships that we assume constrain a trader's pricing at the given snapshot in time and other relationships on which Propositions 1 and 2 depend. We list only the former as axioms and show that the later can be interpreted simply as definitions of conditional expectations, as they do not further constrain a trader's behavior beyond the axioms. Loosely speaking, our axioms are mathematical definitions of rationality and idealizations of ways in which one derivative can be statically replicated by others at a single snapshot in time ("statically" in that there is no assumption that portfolios would later be rebalanced). Second, the interpretation in Section 2.1 of the trader's expectation functional as subjective is not present in Rogers [5] [18]. Yet, there is nothing in the proof of Theorem 1 of Rogers [18] – as there is nothing in the proofs of Propositions 1 and 2 here -- requiring that a trader's expectations are consistent with any model objectively governing the underlying's price process.

In the special case in which a trader does possess knowledge of a model objectively governing the underlying's price process (that is, in which the trader's expectations happen to be correct), Proposition 2 shows that consideration of simple forms of static replication would compel a trader to use prices that are objectively risk neutral. Hence, the mechanics of dynamic portfolio replication are not necessary for market prices to coincide with objectively defined risk neutral prices, as long as the objective knowledge necessary for dynamic portfolio replication is available to traders.





In Section 2.2, we extend the framework of Section 2.1 with more a specific axiom implying that the trader's subjective model for the price of the underlying can be expressed as a time-inhomogeneous diffusion process defined by an SDE. We then apply well known results relating the representation of the risk-neutral pricing functional from Section 2.1 to a second SDE with risk-neutral drift.

In Section 2.3, we introduce an additional axiom stating that the instantaneous variance of returns must exhibit the same time-average behavior over large time scales as does the quadratic variation in (1.1.3), and we show that this axiom is necessarily to prevent arbitrage involving options on the realized quadratic variation. The particular options considered there are closely related to the notion of a variance swap as studied on pages 136-143 of Gatheral [13] and references cited there.

Finally, in Section 2.4, we introduce one final axiom – technical and not constraining in practice – and derive asymptotics for the implied volatilities of European call options as their expiration times become large. Applying the path-from-spot-to-strike result from pages 26-31 of Gatheral [13] (as interpreted by Lee [19]), the results there show that, under the axioms, the implied volatility surface flattens for large expirations. Our result is loosely consistent with asymptotic estimates obtained by Fouque, Panicolaou, and Sincar [19] [20]; see also pages 95-96 of Gatheral [13] for a summary of those results. This flattening of the implied volatility surface is generally seen in empirical studies, as reviewed in Section 3 of Gatheral [13].

The SDE representation for the diffusion in Section 2.2 is a common starting point when modeling the price of a security and is often called the *local volatility model*; see Chapter 1 of Gatheral [13] for background. Dupire [21] showed that risk-neutral pricing obtained from local volatility models can fit any continuum of prices of European call options spanning all possible strike prices and expiration dates. In particular, a local volatility model can fit the *volatility smile* frequently observed for empirical prices of options, as described on pages 37-40 of Gatheral [13]. Breeden and Lizenberger [22] earlier showed how





a continuum of option prices with the same expiration date determine a risk-neutral density as a function of price level. Carr and Madan [23] later showed how to determine the local volatility surface from a continuum of option prices with the same expiration data assuming path independence of the underlying's price. Under weaker assumptions, Carr and Madan [24] showed how to fit a risk-neutral model to a grid of option prices.

As Dupire [25] and Derman and Kani [26] also showed, risk neutral pricing resulting from local volatility models is consistent with pricing under models in which instantaneous variances of returns are themselves quite general stochastic processes. Then the function defining the instantaneous variance of returns for the local volatility model can be interpreted as a conditional expectation of the instantaneous variance for the more general model. Their result supports our contention that a local volatility model is a reasonable idealization of the subjective beliefs of a generic trader. For further properties of the local volatility model, see pages 7-18 and 25-31 of Gatheral [13].

Because the local volatility model possesses those desirable properties, it is commonly used as a tool to insure consistency in pricing of different types of derivatives. Dupire [21] and Derman and Kani [26] fit a local volatility models to a sampling of market prices for options and then use those models for the risk-neutral pricing of other derivatives. In the context of that work, as well as of the work of Breeden and Lizenberger [22] and Carr and Madan [23] [24] described above, the risk neutral model inferred from option prices can be interpreted as subjective, in that it need not accurately reflect the objective price dynamics of the underlying. If such a risk-neutral model did in fact reflect the underlying's price dynamics, then the prices obtained from the model would preclude *all* forms of arbitrage. Therefore, risk-neutral pricing – whether or not an accurate reflection of the underlying's price dynamics – is always *sufficient* (and therefore a good strategy) for precluding certain forms of arbitrage not depending on the model's accuracy, i.e. arbitrage exploiting inconsistencies between the trader's own prices for





different derivatives.  Section 2 of our paper proves the converse that risk-neutral pricing by each trader with respect to his or her subjective model is also *necessary* to prevent such inconsistencies.

The results from Dupire [21] imply that traders using local volatilities models must agree on risk-neutral pricing in general if their models are each calibrated to reproduce the continuum of market prices for European call options. In that case, acceptance of market prices for options by a trader constrains his or her subjective model for the price dynamics of the underlying.  (This is an example of the general phenomena that opinions of different individuals, even though subjective, are typically far from independent.) The results in Section 2.3 show that the subjective local volatility models of traders are similarly constrained -- without accounting for any knowledge of market prices for options -- by the objective knowledge of quadratic variations of the underlying's prices over large time scales. To the extent that local volatility models can be viewed as idealizations of the models used by all traders, the results in Section 2.4 then suggest that the objective knowledge of quadratic variations over large time scales will constrain *market* prices for options with long expirations.

In Section 3, we develop results useful for parameter estimation and for hedging without the benefit of dynamic portfolio replication. We assume there, as in Section 2.2, that a trader models the price dynamics of the underlying via a time-inhomogeneous diffusion process. As Theorem 1.1 on page 165 of Karlin and Taylor [27] shows, a time-inhomogeneous diffusion process is a general representation of a (strong) Markov process with continuous sample paths.  We further assume in Section 3 that centered log returns under the trader's model exhibit the wide-sense Markov property, cf. Definition 2.2  of Mandrekar [28] or pages 90-91 of Doob [29]. The Markov and wide-sense Markov properties are consistent with efficient market hypotheses that future prices should depend on past prices through current prices and that future returns should depend on past returns through current returns. The main insight leading to the results of Section 3 is that any efficient-market argument why log returns should





exhibit Markov properties can be applied equally for incremental log returns, since the increment of a log return is simply the log return relative to a different start time. The assumption that incremental log returns possess the wide-sense Markov property turns out to tightly constrain the characteristics of the underlying's price process.

The results of Section 2.4 suggest that it should be possible to estimate the parameter $\alpha$ defined in (1.1.3) from the implied volatility surface derived from option prices. Nevertheless, for static hedging of the underlying's volatility risk, it may be desirable to estimate the underlying's volatility directly from the underlying's historic prices. While it is possible to estimate $\alpha$ in a model-independent way by directly estimating (1.1.2) for historic returns, doing so would require both the use of high-frequency pricing data and, in practice, some form of filtering to remove effects of market microstructure as described in Barndoff-Nielsen and Shepard [30]. An alternative approach suggested by the analysis here is to assume that past returns followed a time-inhomogeneous diffusion process and use relationships between the parameter $\alpha$ and quantities that can be estimated for that process from past data. Maintaining the viewpoint from Section 2 of a trader at a single snapshot in time, Section 3 extends the time domain of the SDE model from Section 2 into the past to model historic prices. We then show how log-return covariances for the model depend on the parameter $\alpha$ at large time scales. The covariance function looks asymptotically like the covariance function that holds for any continuous Gaussian Markov process with stationary increments, as derived in Fendick [31]. This demonstrates a sense in which (1.1.3) implies asymptotic stationarity of increments of the underlying price process at large time scales.

In the setting of this paper, hedging is the construction of portfolios that use derivatives to offset the underlying's volatility risk. As discussed on pages 121-124 of Ross [32], an approach to hedging that does not require the feasibility of dynamic portfolio replication is to choose from among alternative portfolios based on a present-value analysis of each. Since future payoffs from different derivatives in a portfolio





may occur at different times, the distribution of a portfolio's present value will depend on the finite dimensional distributions of the underlying's price process.  Under the assumptions discussed above  of continuous sample paths and Markov properties for incremental log returns, we derive a limit in Section 3 in which the finite-dimensional distributions of centered log returns depend on the underlying's price process only through the parameter $\alpha$. This result is a multi-dimensional central limit theorem for a time-inhomogeneous diffusion. Central limit theorems have been previously derived for time-homogeneous diffusions; see for example Theorem 9 on page 94 of Mandl [33] and Section 3 of Whitt [34].

The results of Section 3 imply that log returns will always exhibit two characteristics commonly observed in empirical studies of asset prices (cf. Cont [35]):

- *volatility clustering*: autocorrelations of squared returns tend to be positive over a broad range of time scales. High volatilities tend to follow high volatilities, and low volatilities tend to follow low volatilities

- *aggregational normality*: the distribution of log returns looks Gaussian over large time scales but not small ones

These characteristics are frequently used as design constraints for models and justification for their dynamic structure.  The results here show that they are in fact inherent characteristics of continuous processes exhibiting convergent quadratic variations and Markov properties consistent with efficient market assumptions.

Empirical studies by Lo and MacKinlay [36] and Conrad and Kaul  [37] have also concluded that the returns of portfolios such as stock index funds exhibit positive autocorrelations; see Fama [38] and Boudoukh, Richardson, and Whitelaw [39] for subsequent interpretations of those results. The results in Section 3 show that, for continuous processes defined on the half-line, negative autocorrelations are in





fact incompatible with our Markov assumptions. Since the Markov properties that we assume are characteristics of efficient markets, empirical studies showing non-negative autocorrelations of portfolio returns support the hypothesis that markets for portfolios, such as index funds, are efficient over a wide range of time scales. Nevertheless, we also discuss in Section 3.3 how our model can be extended to exhibit negative autocorrelations over bounded time domains. We focus in Section 3.1 and 3.2 on the case in which the time domain is unbounded from above to enable the reader to grasp our results as easily as possible. The allowed parameter ranges in the statements of our theorems would become more complicated in the general case.

## 2    Derivative Pricing

In this section, we study derivative pricing from the perspective of an individual trader with limited objective knowledge about properties of the underlying's price trajectories. We will assume throughout that the risk-free interest rate is a constant $r$, that the underlying pays no dividend, that no taxes or fees apply for payouts or trades, and that bid-ask spreads are negligible.

### 2.1    Risk Neutral Yet Subjective

We begin by studying constraints on derivative pricing required to preempt some specific arbitrage strategies without assuming *any* knowledge by the trader of the objective price dynamics of the underlying. Even without such knowledge, arbitrage is still possible if derivatives are mispriced relative to one another. We formulate a set of axioms constraining derivative pricing at a single snapshot in time precluding certain such forms of arbitrage and show how, under those axioms, derivative pricing used by an individual trader must be risk-neutral with respect to his or her subjective expectations about the present value of the derivatives' potential payouts.





We will assume that, at the fixed snapshot $V \geq 0$ in time, the trader knows the past trajectory of prices of the underlying on the interval $[0, V]$ and has probabilistic expectations about the future trajectory on the bounded interval $(V, V + H)$. We will continue to let $S(t)$ denote the price of the underlying at time $t$, but will let $\{S_V(t) : 0 \leq t \leq V + H\}$ denote the set of possible trajectories of those prices confined to ones with known values on $0 \leq t \leq V$, and we will treat $\Omega \equiv \{S_V(t) : V \leq t \leq V + H\}$ in this section as the sample space for which a trader will assign probabilities. We define the sample space in this way so as not to require that traders make probabilistic assumptions about the prior history of the underlying's price.

For $V \leq t \leq V + H$, let $\mathfrak{F}_t$ denote the sigma field generated by $S_V(\cdot)$ on the interval $[0, t]$. If a derivative pays a random amount $p$ at time $t \in [V, V + H]$, where $p$ is determined by the prices $S_V(\cdot)$ up to time $t$, then the discounted present value of that payout at time $V$ is $e^{-r(t-V)}p$. We will then let $E_V[e^{-r(t-V)}p]$ denote the trader's expectation at time $V$ about the discounted present value of that derivative's future payout. We assume only that $E_V[\cdot]$ can be expressed as an expectation functional with respect to some probability measure specific to the given trader. Formally, we will assume that all payouts are scalar valued $\mathfrak{F}_{V+H}$ measurable functions, that a payout occurring at time $t \in [V, V + H]$ is $\mathfrak{F}_t$ measurable, and that there exists a filtered probability space $(\Omega, \mathfrak{F}_{V+H}, P_V)$ such that the following axiom holds:

[A1]     $E_V[f] = \int f dP_V$ for any $\mathfrak{F}_{V+H}$ measurable $f$ satisfying $\int |f| dP_V < \infty$.

For $V < t \leq V + H$ and the same assumptions about $f$, we will also let $E_t[f]$ denote any $\mathfrak{F}_t$ measurable function satisfying

$$E_V[I(A)E_t[f]] = E_V[I(A)f] \text{ for any } A \in \mathfrak{F}_t \tag{2.1.1}$$

where $I(A) = 1$ when $S_V(\cdot) \in A$, and $I(A) = 0$ otherwise. Then, $E_t[f]$ has the interpretation as the conditional expectation of $f$ relative to $\mathfrak{F}_t$. It is well known that conditional expectations are unique up





to null sets; (cf. page 466 of Billingsley [40]). For now, we will assume that $E_t[\cdot]$ is implicitly defined

through (2.1.1) by the trader's unconstrained choice of any expectation functional $E_V[\cdot]$ satisfying [A1].

A conditional expectation satisfying (2.1.1) will always exists, as proven on page 466 of Billingsley [40],

so that (2.1.1) is not itself an assumption about the trader's behavior. In Section 2.2, we will introduce

an additional axiom formalizing a trader's beliefs about the conditional expectation functional and

hence further constraining $E_V[\cdot]$.

We next assume the existence of a scalar-valued functional $\breve{E}_V[\cdot]$ defining the trader's price at time $V$ for

any given derivative (i.e., the maximum price at which the trader would buy it and the minimum price at

which the trader would sell it) based on the present value of its payout. For a derivate paying a

(random) amount $p$ at time $t \in [V, V + H]$ determined by the prices $S_V(\cdot)$ up to time $t$, we will denote

the trader's price at time $V$ by $\breve{E}_V[e^{-r(t-V)}p]$. We will assume that the derivative pricing functional $\breve{E}_V[\cdot]$

has same domain as does the expectation functional $E_V[\cdot]$ and satisfies the following axioms:

[A2]     $\left|\breve{E}_V[f]\right| \leq M E_V[|f|]$ *for some constant $M$ not depending on $f$.*

[A3]     $\breve{E}_V[a_1 f_1 + a_2 f_2] = a_1\breve{E}_V[f_1] + a_2\breve{E}_V[f_2]$ *whenever $a_1$ and $a_2$ are constants.*

[A4]     $\breve{E}_V[f] \geq 0$ *whenever $f \geq 0$.*

[A5]     $\breve{E}_V[1] = 1$.

[A6]     $\breve{E}_V[f] = \lim \breve{E}_V[f_n]$ *if $f_n$ increases monotonically to $f$.*

[A7]     If $f \geq 0$, *then $\breve{E}_V[f] = 0$ if and only if $P_V(f > 0) = 0$.*

[A8]     $\breve{E}_V\left[e^{-r(t_1-V)} I(A) S_V(t_1)\right] = \breve{E}_V\left[e^{-r(t_2-V)} I(A) S_V(t_2)\right]$ *for any $V \leq t_1 \leq t_2 \leq V + H$ and*

      *any $A \in \mathfrak{F}_{t_1}$.*

Axioms [A3] and [A7] are conditions that any rational agent can be expected to observe. Axiom [A2]

requires only that $M$ is some finite value. In practice, we would expect [A2] to hold when $M = 1$

because a trader purchasing a derivative will demand a risk premium. Axiom [A7] says that a trader will





pay a positive amount for a derivative only if he or she believes there is some chance its payout will be positive, and will give a derivative away for free only if he or she believes there is no such chance. Axiom [A7] precludes arbitrage in cases in which a trader's assessment of the impossibility of a positive payout is in fact objectively correct; see the proof of Proposition 4 for an example of such a scenario. The other axioms define conditions that, if violated, would present counterparties with opportunities for certain gain with no risk. Deriving a strategy to capitalize on violations to each such condition is straightforward. We provide examples for axiom [A8] at the end of this section.

Analogously to (2.1.1), we will define $\breve{E}_t[\cdot]$ for each $t \in (V, V+H]$ by the properties that, for any $\mathfrak{F}_{V+H}$ measurable function $f$ satisfying $\int |f| dP_V < \infty$, $\breve{E}_t[f]$ is $\mathfrak{F}_t$ measurable and

$$\breve{E}_V\left[I(A)\breve{E}_t[f]\right] = \breve{E}_V[I(A)f] \text{ for any } A \in \mathfrak{F}_t. \tag{2.1.2}$$

The following proposition, which depends on [A1]-[A7] but not on [A8], generalizes Theorem 1 of Rogers [18] for payouts with no fixed upper bound. Close analogs to axioms [A1]-[A7] are found in the statement of Theorem 1 of Rogers [18]. Although Theorem 1 of Rogers [18] also includes a close analog to (2.1.2) as an axiom, the proof below of our proposition shows that $\breve{E}_t[f]$ satisfying (2.1.2) will always exist under axioms [A1]-[A7], so that (2.1.2) does not itself constrain the trader's choice of the pricing operator $\breve{E}_V[\cdot]$. Therefore, (2.1.1) and (2.1.2) can be regarded simply as mathematical definitions – and Proposition 1 as a mathematical relationship between them -- about which a trader need not be cognizant when making pricing at time $V$.

*Proposition 1: If (i) $E_V$ is an expectation functional satisfying [A1], (ii) $E_t$ for each $V < t \leq V+H$ is defined by (2.1.1), (iii) $\breve{E}_V$ satisfies axioms [A2]-[A7], and (iv) $\breve{E}_t$ for each $V < t \leq V+H$ is defined by (2.1.2), then $\breve{E}_V$ is also an expectation functional, $\breve{E}_t$ for each $V < t \leq V+H$ is the associated conditional expectation functional relative to $\mathfrak{F}_t$, and there is a strictly positive process $\{g_t : V \leq t \leq V+H\}$ for which*





$$\breve{E}_{t_1}[f] = E_{t_1}[fg_{t_2}]/g_{t_1} \text{ whenever } V \leq t_1 \leq t_2 \leq V + H \tag{2.1.3}$$

for any $\mathfrak{F}_{t_2}$ measureable function $f$ for which $\int |f| dP_V < \infty$. Moreover, $g_t = E_t[g_{V+H}]$ for all $V \leq t \leq V + H$.

*Proof:* By [A2] and [A3], $\breve{E}_V$ is a bounded linear functional defined for $\mathfrak{F}_{V+H}$ functions $f$ for which $\int |f| dP_V < \infty$. Hence, for each $t \in [V, V + H]$, $\breve{E}_V$ is also a bounded linear functional when restricted to $\mathfrak{F}_t$ measureable functions. Since $P_V$ is a probability measure, it is sigma-finite by definition. It then follows immediately from the Riesz Representation Theorem on page 246 in Chapter 11, Section 7 of Royden [41] that for each $t \in [V, V + H]$ there is a bounded $\mathfrak{F}_t$ measureable function $g_t$, unique except on null sets of $P_V$, such that

$$\breve{E}_V[f] = \int f g_t \, dP_V = E_V[fg_t] \tag{2.1.4}$$

for any $\mathfrak{F}_t$ measureable function $f$ for which $\int |f| dP_V < \infty$.

For any $A \in \mathfrak{F}_{V+H}$, let

$$\breve{P}_V(A) \equiv \breve{E}_V[I(A)] . \tag{2.1.5}$$

By [A3]-[A6], $\breve{P}_V$ is a probability measure (cf. Theorem 3.2.1 on page 42 of Whittle [42]), $\breve{E}_V[\cdot]$ from (2.1.4) is an expectation functional; and $\breve{E}_t[\cdot]$ from (2.1.2) is the corresponding conditional expectation functional, which will always exist (cf. page 466 of Billingsley [40]). Using [A7], $\breve{P}_V(A) = 0 \Leftrightarrow \breve{E}_V[I(A)] = 0 \Leftrightarrow P_V[I(A) > 0] = 0 \Leftrightarrow P_V[A] = 0$, so that $P_V$ and $\breve{P}_V$ are absolutely continuous with respect to one another. Since $\breve{P}_V$ is absolutely continuous with respect to $P_V$ and $\breve{P}_V(A) = \int I(A) g_t \, dP_V$ for any $A \in \mathfrak{F}_t$ by (2.1.4) and (2.1.5), the function $g_t$ in (2.1.4) must be non-negative almost surely for $V \leq t \leq V + H$, as follows from the uniqueness of $g_t$ and the Radon-Nikodym theorem (cf. Theorem 23 on page 238 of Royden [41]). Since $P_V$ is absolutely continuous with respect to $\breve{P}_V$, the function $g_t^{-1}$ must be finite





with probability one under $P_V$, as also follows from the Randon-Nikodym theorem, so that $g_t$ can be taken to be strictly positive.

If $f$ is a constant, then $E_V[f] = f = \breve{E}_V[f]$ where the second equality follows from axioms [A3] and [A5]. This implies that $g_V = 1$ by (2.1.4), so that (2.1.3) follows from (2.1.4) for the case in which $V = t_1 \leq t_2 \leq V + H$. By [A5] and (2.1.4), $E_V[g_{V+H}] = 1 = g_V$.

If $A \in \mathfrak{F}_{t_1}$ and $f$ is $\mathfrak{F}_{t_2}$ measurable for $V < t_1 \leq t_2 \leq V + H$, then $\breve{E}_{t_1}[f]$ is $\mathfrak{F}_{t_1}$ measurable by its definition, and

$$E_V[I(A)\,\breve{E}_{t_1}[f]g_{t_1}] = \breve{E}_V\left[I(A)\breve{E}_{t_1}[f]\right] = \breve{E}_V[I(A)f] = E_V[I(A)fg_{t_2}] = E_V\left[I(A)\,E_{t_1}[fg_{t_2}]\right] \qquad (2.1.6)$$

where the first equality follows by (2.1.4), the second from (2.1.2), the third from (2.1.4) again, and the last from (2.1.1). Since conditional expectations are unique up to null sets, the equality of the first and last expressions of (2.1.6) implies that

$$\breve{E}_{t_1}[f]g_{t_1} = E_{t_1}[fg_{t_2}]. \qquad (2.1.7)$$

Since $g_{t_1} > 0$, (2.1.3) follows for the case in which $V < t_1 \leq t_2 \leq V + H$. Setting $t_1 = t$, $t_2 = V + H$ and $f = 1$ in (2.1.7) and applying [A5] shows that $g_t = E_t[g_{V+H}]$ for $V < t \leq V + H$. $\blacksquare$

Since $\breve{P}_V(A) = \int I(A)\, d\breve{P}_V = \int I(A)g_{V+H}\, dP_V$ for any $A \in \mathfrak{F}_{V+H}$, the function $g_{V+H} = d\breve{P}_V/dP_V$ is a Radon-Nikodym derivative defining a change of measure, and $g_t = E_t[g_{V+H}] = E_t[d\breve{P}_V/dP_V]$ for $V \leq t \leq V + H$ is a density process (using the terminology from Section F, Chaper 6 of Duffie [7]). Two probability measures are equivalent if they have the same null sets. As the proof of Proposition 1 shows, $P_V$ and $\breve{P}_V$ are equivalent measures.

*Proposition 2: If [A1]- [A8] hold, then, in addition to the conclusions of Proposition 1,*

$$\breve{E}_{t_1}\left[e^{-r(t_2-V)}S_V(t_2)\right] = e^{-r(t_1-V)}S_V(t_1) \text{ whenever } V \leq t_1 \leq t_2 \leq V + H.$$





*Proof :* If $V = t_1 \leq t_2 \leq V + H$, the conclusion of Proposition 2 follows immediately from [A8].

Otherwise, when $V < t_1 \leq t_2 \leq V + H$ and $A \in \mathfrak{F}_{t_1}$,

$$\breve{E}_V \left[ I(A) \breve{E}_{t_1} [ e^{-r(t_2 - V)} S_V(t_2) ] \right] = \breve{E}_V \left[ I(A) e^{-r(t_2 - V)} S_V(t_2) \right]$$

$$= \breve{E}_V \left[ I(A) e^{-r(t_1 - V)} S_V(t_1) \right] \qquad (2.1.8)$$

where the first equality follows from the definition (2.1.2) and the second from [A8]. By Proposition 1, $\breve{E}_t[\cdot]$ for any $t \in (V, V + H]$ is a conditional expectation functional. Since conditional expectations are unique up to null sets, (2.1.8) implies that $\breve{E}_{t_1}[ e^{-r(t_2 - V)} S_V(t_2)] = e^{-r(t_1 - V)} S_V(t_1)$. ∎

Proposition 2 says that the discounted security price is necessarily a martingale under the trader's subjective pricing measure. We can therefore paraphrase the conclusion of Proposition 2 by saying that $\breve{P}_V$ is *an equivalent martingale measure* with respect to the subjective probability measure $P_V$ or more succinctly that $\breve{P}_V$ is a *risk-neutral measure* relative to $P_V$.

Axiom [A8] is the only axiom reflecting the assumption that the risk-free interest rate is a constant $r$. In the statement of [A8], the presence of the indicator function $I(A)$ for condition $A \in \mathfrak{F}_{t_1}$ models the down-and-out scenario in which the derivative expires early with no payout if the security price violates that condition by time $t_1$. If a first derivative pays $I(A)S_V(t_1)$ at time $t_1 \geq V$ and that amount is immediately reinvested in the underlying, then the investment will always yield $I(A)S_V(t_2)$ at time $t_2 \geq t_1$. Axiom [A8] says that a second derivative paying $I(A)S_V(t_2)$ at time $t_2$ should therefore trade at time $V$ at the same price as the first. (Otherwise, a counterparty can achieve a guaranteed profit by selling the more expensive one and buying the less expensive one.) One interpretation of [A8] is that the arbitrage free price at any snapshot in time for the underlying itself is its market price then.

As a special case of [A8],





$$\breve{E}_V\left[e^{-r(t-V)}(S_V(t) - C)\right] = S_V(V) - Ce^{-r(t-V)} \text{ for any } V \leq t \leq V + H. \tag{2.1.9}$$

The term $S_V(t) - C$ on the left-hand side can be interpreted as the payout of a forward contract on the

underlying with delivery price $C$ at expiration time $t$. The right-hand side of (2.1.9) is the unique price at

time $V$ of such a forward contract preventing arbitrage (a well-known result); see for example 5.2b on

page 67 of Ross [32]. If

$$K^+ \equiv \begin{cases} K, & \text{if } K > 0 \\ 0, & \text{otherwise}, \end{cases}$$

then a *European call option* with expiration time $t \in [V, V + H]$ and exercise price $C$ is a derivative that

pays $(S_V(t) - C)^+$ at time $t$. A European put option with the same expiration time and exercise price is

a derivative that pays $\left(C - S_V(t)\right)^+$ at time $t$. Because $(x - C)^+ + (C - x)^+ = x - C$ for any real

number $x$, the left-hand side of (2.1.9) also has the interpretation as the trader's price for a portfolio

that is long one European call option and short one European put option, and the equality of (2.1.9) is

the well known formula for put-call parity at time $V$ ; see pages 66-68 of Ross and Section 4.5.6 on pages

162-164 of Shreve [6] for further background on those formulas.

## 2.2   Conditional Expectations

The properties described by Propositions 1 and 2 do not uniquely define derivative prices when the

subjective measure $P_V$ is arbitrary, but they do under more specific assumptions about the trader's

model of the underlying's price dynamics.  To make further progress, we idealize the trader's subjective

model for those dynamics through an additional axiom:

[A9]    *$P_V$ is the probability measure for the solution to*

$$\frac{dS_V(t)}{S_V(t)} = \mu_V(S_V(t), t)dt + \sigma_V(S_V(t), t)dB(t) \text{ for } V \leq t \leq V + H$$

*when $S_V(V)$ is known and $B$ is some Brownian motion.*





Axioms [A9] defines a diffusion process. The functions $\mu_V(\cdot,\cdot)$ and $\sigma_V(\cdot,\cdot)^2$ are commonly called the diffusion coefficients, and $\sigma_V(\cdot,\cdot)$ the volatility surface. The results that follow do not require that the trader bases pricing decision at times other than $V$ on the same model.

Axiom [A9] implies that the trader at the particular snapshot $V$ in time believes that price trajectories $S_V$ will be continuous and that the mean and variance of the instantaneous returns will be described, respectively, by

$$E_t[dS_V(t)/S_V(t)] = \mu_V(S_V(t),t)dt \quad and \quad E_t[dS_V(t)^2/S_V(t)^2] = \sigma_V(S_V(t),t)^2dt, \qquad (2.2.1)$$

for $V \leq t \leq V + H$ where $dS_V(t) \equiv S_V(t+dt) - S_V(t)$. Conversely, Theorem 3.3 on pages 287-288 of Doob [29] shows that [A9] follows from continuity of sample paths and (2.2.1). In other words, a trader need not think in terms of Brownian motion for [A9] to be a reasonable idealization of his or her probabilistic expectations about the future. Doob's result assumes regulatory conditions for the diffusion coefficients that are implicit in [A9]. Page 288 of Doob [29] describes additional regulatory conditions for the diffusion coefficients, also implicit in [A9], under which the solution to the SDE is unique.

When (2.2.1) holds, we would expect from Proposition 2 that $\breve{E}_t[dS_V(t)/S_V(t)] = rdt$ and from Proposition 1 that $\breve{E}_t[dS_V(t)^2/S_V(t)^2]/E_t[dS_V(t)^2/S_V(t)^2] \to 1$ as $dt \to 0$. (To see why the later should hold, note that $\breve{E}_t[dS_V(t)^2/S_V(t)^2] = E_t[(dS_V(t)^2/S_V(t)^2)(g_{t+dt}/g_t)]$ by (2.1.3) and that $E_t[dS_V(t)^2/S_V(t)^2] = \breve{E}_t[(dS_V(t)^2/S_V(t)^2)(g_{t+dt}^{-1}/g_t^{-1})]$ since $\breve{P}_V$ and $\breve{P}_V$ are equivalent measures. The asymptotic equivalence of $\breve{E}_t[dS_V(t)^2/S_V(t)^2]$ and $E_t[dS_V(t)^2/S_V(t)^2]$ then follows by applying Holder's inequality to the right hand side of each of those equations and taking appropriate limits assuming that $\{g_t\}$ is continuous.) The following proposition confirms that these properties hold.





Proposition 3: *If axioms [A1]-[A9] hold*, then $\breve{P}_V$ *defined in* (2.1.5) *is the probability measure for the solution to*

$$\frac{dS_V(t)}{S_V(t)} = r\,dt + \sigma_V(S_V(t),t)dB(t) \; for \; V \le t \le V + H \tag{2.2.2}$$

*when* $S_V(V)$ *is known and B is some Brownian motion.*

For a formal proof of Proposition 3 utilizing the properties described by Propositions 1 and 2 from Section 2.1, see Chapter 6 of Duffie [7]. ∎

## 2.3 Time Average Variances

The rational for our next axiom is provided by the proposition that follows it:

[A10]   *For some* $\alpha > 0, \theta > 0, \; \gamma > 0, and \; T_0 \in (0,H]$,

$$P_V\left(\left|\frac{1}{T}\int_V^{V+T} \sigma_V(S_V(w),w)^2 \, dw - \alpha\right| \le \frac{\theta}{T^\gamma}\right) = 1 \; on \; T_0 \le T \le H \;.$$

Proposition 4: *If axioms [A1]-[A9] hold and prices* $\{S(t): t \ge 0\}$ *of the underlying have the property defined by (1.1.3) for some* $T_0 \in (0,H]$ *, then* [A10] *is also necessary to prevent arbitrage.*

*Proof:* Let

$$X_V^{(0)}(t) \equiv log(S_V(t)/S_V(0)) \; \text{ for } 0 \le t \le V + H. \tag{2.3.1}$$

By Ito's lemma and Proposition 3,

$$dX_V^{(0)}(t) = \left(r - \frac{1}{2}\sigma_V(S_V(t),t)^2\right)dt + \sigma_V(S_V(t),t)dB(t) \tag{2.3.2}$$

for $V \le t \le V + H$ under $\breve{P}_V$. Lemma 4.4.4 on page 143 of Shrive [6] then implies that





$$\check{P}_V \left( \left[ X_V^{(0)} \right]_{V,V+T} = \int\limits_V^{V+T} \sigma_V(S_V(t),t)^2 \, dt \right) = 1$$

for $0 < T \leq H$. This, in turn, implies that

$$\check{E}_V \left[ e^{-rT} \left( \frac{1}{T} \left[ X_V^{(0)} \right]_{V,V+T} - \alpha - \frac{\theta}{T^\gamma} \right)^+ \right] = \check{E}_V \left[ e^{-rT} \left( \frac{1}{T} \int\limits_V^{V+T} \sigma_V(S_V(t),t)^2 \, dt - \alpha - \frac{\theta}{T^\gamma} \right)^+ \right]. \qquad (2.3.3)$$

for any $\alpha, \theta$, and $\gamma$. The left hand side of (2.3.3) has the interpretation as the trader's price at time $V$ of a European call option defined for the time-average quadratic variation of $X_V^{(0)}$ over $[V, V+T]$ when the strike price is $\alpha + \frac{\theta}{T^\gamma}$ and expiration time is $V + T$.

Since (1.1.3) is assumed to hold for all sample paths, it must hold in particular for the constrained samples paths $X_V^{(0)}$. When $\alpha, \theta$, and $\gamma$ are the parameters from (1.1.3) and $T_0 \in (0, H]$, it then follows from (1.1.3) that

$$\frac{1}{T} \left[ X_V^{(0)} \right]_{V,V+T} \leq \alpha + \frac{\theta}{T^\gamma} \ on \ \ T_0 \leq T \leq H \ .$$

For such values of parameters, the left-hand side of (2.3.3) must equal zero, else a profit would result with probability one from selling such an option at time $V$. (This is an example where axiom [A7] prevents arbitrage.) But then the equality in (2.3.3) implies that $\frac{1}{T} \int_V^{V+T} \sigma_V(S(t),t)^2 \, dt \leq \alpha + \frac{\theta}{T^\gamma}$ with probability one under $\check{P}_V$. A similar argument involving a European put option for the time-average quadratic variation shows that $\frac{1}{T} \int_V^{V+T} \sigma_V(S(t),t)^2 \, dt \geq \alpha - \frac{\theta}{T^\gamma}$ with probability one under $\check{P}_V$. Because $\check{P}_V$ and $P_V$ have the same null sets, we conclude that [A10] is necessary to avoid arbitrage. ∎

## 2.4   Implied Volatilities

The conditions of axiom [A10] are satisfied if





$$\frac{1}{T}\int\limits_{V}^{V+T} \sup_{s\geq 0}\sigma_V(s,t)^2\, dt \leq \alpha + \frac{\theta}{T^\gamma} \quad and \quad \frac{1}{T}\int\limits_{V}^{V+T}\inf_{s\geq 0}\sigma_V(s,t)^2\, dt \geq \alpha - \frac{\theta}{T^\gamma} \qquad (2.4.1)$$

for $T_0 \leq T \leq H$. It is not necessarily true that (2.4.1) holds whenever [A10] does, since there need not exist continuous trajectories $S_V(\cdot)$ such that $\sigma_V(S_V(t),t)^2$ is equal to $\sup_{s\geq 0}\sigma_V(s,t)^2$ or to $\inf_{s\geq 0}\sigma_V(s,t)^2$ for all $V \leq t \leq V + T$. But we lose little generality by assuming the following (final) axiom:

[A11]   $\sigma_V(\cdot,\cdot)^2$ is uniformly Lipschitz in its first argument and bounded from above, and there exist functions $s_1(\cdot)$ and $s_2(\cdot)$, differentiable almost everywhere on $[V, V + H]$, for which

$$\sigma_V(s_1(t),t)^2 = \inf_{s\geq 0}\sigma_V(s,t)^2 \text{ and } \sigma_V(s_2(t),t)^2 = \sup_{s\geq 0}\sigma_V(s,t)^2 \text{ for all } V \leq t \leq V + H.$$

A uniform Lipschitz condition is already implicit in [A9] (see for example Page 288 of Doob [29]), and the assumed upper bound on $\sigma_V(\cdot,\cdot)^2$ can be arbitrarily large. Functions $s_1(\cdot)$ and $s_2(\cdot)$ satisfying [A11] need not be continuous. For example, a sufficient condition for a function to be differentiable almost everywhere on a closed bounded interval is that it can be represented there as the difference of two monotone functions (cf. Corollary 5 on page 100 of Royden [41]).

Lemma 1: *If axioms [A9]-[A11] hold, then (2.4.1) holds for parameters constrained as in [A10].*

*Proof*: Let $q$ denote Lebesgue measure on $[V, V + H]$. Since $s_2(\cdot)$ defined in [A11] is differentiable almost everywhere, there exists, for any $0 < \delta \leq H$, a continuously differentiable function $s_\delta(\cdot)$ such that $q(\{t: s_2(t) \neq s_\delta(t)\}) < \delta$ and $s_\delta(V) = S_V(V)$. (By Theorem 1 of Whitney [43], there exists a continuously differentiable function $s_\delta(\cdot)$ such that $q(\{t: s_2(t) \neq s_\delta(t)\}) < \delta/2$, and it can be modified on $V \leq t \leq V + \delta/2$ so that it remains continuously differentiable and satisfies $s_\delta(V) = S_V(V)$. ) By [A11], $\sigma_V(\cdot,\cdot)^2 \leq M$ for some $M > 0$, so that





$$\left| \int\limits_{V}^{V+T} \sigma_V(s_2(t),t)^2 dt - \int\limits_{V}^{V+T} \sigma_V(s_\delta(t),t)^2 \, dt \right| \le M\delta. \tag{2.4.2}$$

For given $\varepsilon > 0$, let $\mathcal{S}_{\varepsilon,\delta}$ denote the set of continuous trajectories $S_V(\cdot)$ such that $|S_V(t) - s_\delta(t)| < \varepsilon$ on $V \le t \le V + H$ . When $S_V(\cdot) \in \mathcal{S}_{\varepsilon,\delta}$,

$$|\sigma_V(S_V(t),t)^2 - \sigma_V(s_\delta(t),t)^2| \le K|S_V(t) - s_\delta(t)| \le K\varepsilon$$

on $V \le t \le V + H$ for some some $K \ge 0$ not depending on the arguments of $\sigma_V(\cdot,\cdot)^2$, as is implied by the uniform Lipschitz condition from [A11]. Therefore,

$$\left| \int\limits_{V}^{V+T} \sigma_V(S_V(t),t)^2 dt - \int\limits_{V}^{V+T} \sigma_V(s_\delta(t),t)^2 \, dt \right| \le K\varepsilon T. \tag{2.4.3}$$

when $S_V(\cdot) \in \mathcal{S}_{\varepsilon,\delta}$ By (2.4.2), (2.4.3), and the triangle inequality,

$$\left| \int\limits_{V}^{V+T} \sigma_V(S_V(t),t)^2 dt - \int\limits_{V}^{V+T} \sup_{s \ge 0} \sigma_V(s,t)^2 \, dt \right| = \left| \int\limits_{V}^{V+T} \sigma_V(S_V(t),t)^2 dt - \int\limits_{V}^{V+T} \sigma_V(s_2(t),t)^2 \, dt \right|$$

$$\le K\varepsilon T + M\delta \tag{2.4.4}$$

when $S_V(\cdot) \in \mathcal{S}_{\varepsilon,\delta}$.

As Maruyama [44] first showed, the probability measure $P_V$ for a time-inhomogeneous diffusion as in [A9] has the same null sets as does the probability measure of a Brownian motion $B$ with the same starting point, so that $S_v - s_\delta$ under $P_V$ has the same null sets as does $B - s_\delta$. In turn, the probability measure for $B - s_\delta$ on $[V, V + H]$ has the same null sets as does the probability measure of a Brownian motion $B^*$ with zero drift starting at zero, as follows from the example of Girsanov's transformation at the top of page 198 of Chung and Williams [45] since, on that closed interval, the derivative $s'_\delta(\cdot)$ is bounded (being continuous there), and $s_\delta(t) = s_\delta(V) + \int_V^{V+t} s'_\delta(w) \, dw$ by Theorem 7.21 on page 149 of Rudin [46]. The distribution for the exit time of a Brownian motion $B^*$ with zero drift from $[-\varepsilon, \varepsilon]$ is well known not to have finite support (cf., example 6 of Kahale [47]), so we conclude that





$$P_V\big(\mathcal{S}_{\varepsilon,\delta}\big) > 0. \qquad (2.4.5)$$

We then conclude from (2.4.4) and (2.4.5) that

$$ess\ sup \int_V^{V+T} \sigma_V(S_V(t),t)^2\, dt = \int_V^{V+T} \sup_{s\geq 0} \sigma_V(s,t)^2\, dt \qquad (2.4.6)$$

where $ess\ sup\ f \equiv inf\{\tau : P_V(f > \tau) = 0\}$. On the other hand, [A10] implies that

$P_V\left(\int_V^{V+T} \sigma_V(S_V(t),t)^2\, dt > T\left(\alpha + \frac{\theta}{T^\gamma}\right)\right) = 0$, in contradiction to (2.4.6) unless the first equality of

(2.4.1) holds. The second equality of (2.4.1) holds by similar logic. ∎

For $0 \leq t \leq H$, recall that $\breve{E}_V[e^{-rt}(S_V(V+t) - C)^+]$ defines the trader's price at time $V$ for a

European call option with strike price $C$ and expiration time $V + t$. The Black-Sholes implied volatility

$\sigma_V^{BS}(V+t, C)$ for $0 \leq t \leq H$ can then be defined as the unique value for which that option price

$\breve{E}_V[e^{-rt}(S_V(V+t) - C)^+]$ is obtained by taking the expectation of $e^{-rt}(S_V(V+t) - C)^+$ assuming

that $\log S_V(V+t)/S_V(V)$ is a normally distributed random variable with mean $rt - \sigma_V^{BS}(V+t,C)^2 t/2$

and variance $\sigma_V^{BS}(V+t,C)^2 t$; see Chapter 7 of Ross [32] for background on the Black-Sholes formula

and implied volatility.

Proposition 5: *If axioms [A1]-[A11] hold, then*

$$\sigma_V^{BS}(V+T, C)^2 = \alpha\left(1 + O\left(\frac{1}{T}\right)^\gamma\right) on\ T_0 < T \leq H.$$

Proof: By (2.9) and (2.10) of Lee [48],

$$\sigma_V^{BS}(V+T, C)^2 = \frac{1}{T}\int_V^{V+T} \int_0^\infty \sigma_V(s,t)^2\, k_{S(V),C}(s,t) ds\, dt \qquad (2.4.7)$$

where $k_{S,C}(\cdot,\cdot)$ is a positive function depending on $S > 0$ and $C > 0$ and satisfying





$$\int_0^\infty k_{S,C}(s,t)ds = 1 \; for \; each \; V \leq t \leq V + T;$$
(2.4.8)

see also pages 26-31 of Gatheral [13] for related discussion. By (2.4.7) and (2.4.8),

$$\frac{1}{T}\int_V^{V+T} \inf_{s \geq 0} \sigma_V(s,t)^2 \, dt \leq \sigma_V^{BS}(V+T,C)^2 \leq \frac{1}{T}\int_V^{V+T} \sup_{s \geq 0} \sigma_V(s,t)^2 \, dt.$$
(2.4.9)

(Alternatively, (2.4.9) follows directly from Theorem 8 of Bergman, Grundy, and Wiener [49] and the "No Skew" representation of the implied volatility on pages 13 of Gatheral [13].) The statement of Proposition 5 follows immediate from (2.4.9) by Lemma 1 and the definition of a big-oh estimate from Section 1.1 ∎

# 3   Present-value Analysis

We next derive limit theorems and approximations expressing the finite-dimensional distribution and covariance structure of the underlying's price process in terms of the parameter $\alpha$ from (1.1.3). Since the payouts of derivatives are functions of the underlying's price trajectories, the finite-dimensional distributions of those trajectories determine the distributions of the present value not only of shares of the underlying but also of derivatives with different expiration times. The results here also point the way towards estimating the parameter $\alpha$ appearing in (1.1.3) using covariances associated with the underlying's price process over large time scales.

The results of this section are based on essentially the same diffusion model as studied in Section 2 but under the additional assumption that incremental log returns of the underlying have the *wide-sense Markov property*. In Section 3.1, we present asymptotic results that are most useful for present-value analysis and parameter estimation. In Section 3.2, we derive a canonical representation of the underlying's log return process that is valid for all time scale, and we use it to prove the asymptotic





results from Section 3.1. Section 3.3 concludes the paper with some observations about autocorrelations.

## 3.1 Asymptotics

We begin by extending the time domain of the SDE model from [A9] in Section 2.2 by assuming that at time $V$ the trader models the underlying's price $S$ as satisfying

$$\frac{dS(t)}{S(t)} = \mu_V(S(t), t)dt + \sigma_V(S(t), t)dB(t) \tag{3.1.1}$$

for $0 \leq t < \infty$, where $S(0)$ is assumed known and $\mu_V(\cdot, t)$ and $\sigma_V(\cdot, t) > 0$ agree with the functions of the same names from [A9] for $V \leq t \leq V + H$. We will let $P$ denote the corresponding probability measure for the trajectories of $S$ under this model and $E[\cdot]$ denote the expectation functional for functions of those trajectories. Applying the results of page 282 of Doob [29], $P[\cdot \mid S(V) = S_V(V)]$ is given by the distribution of the solution to the SDE (3.1.1) on $V \leq t \leq V + H$ when $S(V) = S_V(V)$. Because the probability measure for an SDE's solution is uniquely determined by the SDE's coefficients, $P[\cdot \mid S(V) = S_V(V)] = P_V[\cdot]$ when $P_V$ is defined by the SDE from [A9].

For $S$ satisfying (3.1.1), $z \geq 0$, $t \geq 0$, let

$$\lambda^{(z)}(t) \equiv E \, log(S(z+t)/S(z)) = E \int_z^{z+t} \mu_V(S(w), w)dw - \frac{1}{2}E \int_z^{z+t} \sigma_V(S_V(w), w)^2 dw, \tag{3.1.2}$$

as follows from (3.1.1) using Ito's formula. Next, let

$$\bar{X}^{(z)}(t) \equiv \log \frac{S(t+z)}{S(z)} - \lambda^{(z)}(t) \tag{3.1.3}$$

denote the centered log-return at time $t + z$ relative to the security price at time $z$. Note that $\bar{X}^{(z)}$ is a zero-mean process, but not necessarily a martingale. Let





$$r^{(z)}(t_1, t_2) \equiv E\bar{X}^{(z)}(t_1)\bar{X}^{(z)}(t_2) \text{ for } 0 \leq t_1 \leq t_2 \qquad (3.1.4)$$

denote its covariance function.

For $z \geq 0$ and $w \geq 0$, let $\hat{E}_w^{(z)}[\cdot]$ denote the functional that projects a real-valued random function of $\bar{X}^{(z)}$ to the element closest to it according to the mean-square metric in the closed linear manifold generated by $\{\bar{X}^{(z)}(t): 0 \leq t \leq w\}$. In other words, $\hat{E}_w^{(z)}[f]$ is the best linear predictor of $f$ conditional on $\bar{X}^{(z)}$ up to time $w$. As described on page 155 of Doob [29], $\hat{E}_w^{(z)}[\cdot]$ has many of the same properties as a conditional expectation functional including linearity.

We will say that the process $\bar{X}^{(z)}$ has the *wide-sense Markov property* if there exists a function $a^{(z)}(\cdot, \cdot)$ such that

$$\hat{E}_s^{(z)}[\bar{X}^{(z)}(t)] = a^{(z)}(s, t)\bar{X}^{(z)}(s) \text{ for all } 0 \leq s \leq t. \qquad (3.1.5)$$

The wide-sense Markov property means that the best linear predictor of the state of $\bar{X}^{(z)}$ at a particular time given any collection of observations of its state at earlier times is equal to the best linear predictor given the most recent of these observations. The Markov and wide-sense Markov properties are known to be equivalent for Gaussian process but are distinct conditions more generally.

*Proposition 6: If*

i) *(continuity of sample paths and Markov property) the pricing process $S$ satisfies (3.1.1) on $[0, \infty)$ when $S(0)$ is known*

ii) *(wide-sense Markov property) for every $z \geq 0$, the process $\bar{X}^{(z)}$ in (3.1.3) satisfies (3.1.5) for some function $a^{(z)}(\cdot, \cdot)$ that is continuously differentiable*

iii) *(convergence of time-average instantaneous variances) there exist $\alpha > 0$ and $\gamma > 0$ such that*





$$P\left(\frac{1}{T}\int\limits_{z}^{z+T}\sigma_V(S(w),w)^2\,dw = \alpha\left(1+O\left(\frac{1}{T}\right)^\gamma\right)\right)=1$$

*for any $z \geq 0$ and all $T$ sufficiently large ,*

then there exists a non-negative constant $\beta$ such that the covariance function $r^{(z)}(\cdot,\cdot)$ defined in (3.1.4) satisfies

$$r^{(z)}(t_1T,t_2T)=\begin{cases}t_1T(\alpha+\beta t_2T)+O(T^{2-\gamma}), & \gamma>1 \text{ and } \beta>0 \\[2mm] \beta t_1 t_2 T^2 + O(T^{2-\gamma}), & 0<\gamma\leq 1 \text{ and } \beta>0 \\[2mm] \alpha t_1 T + O(T^{1-\gamma}), & \gamma>0 \text{ and } \beta=0\end{cases}\qquad(3.1.6)$$

for $z \geq 0$, $0 < t_1 \leq t_2$, and sufficiently large $T$.

Conditions (i) and (ii) of Proposition 6 are consistent with a model in which markets are efficient. When condition (i) above holds, condition (iii) is required for consistency with (1.1.4), as follows from Lemma 4.4.4 of Shreve [6].

For random variables $X$ and $Y$, let $p(X,Y) \equiv Cov(X,Y)/\left(\sqrt{Var\,X}\sqrt{Var\,Y}\right)$ denote their correlation coefficient. As is well known, $-1 \leq p(X,Y) \leq 1$ whenever it is finite. For $X^{(z)}$ defined in (3.1.3), also let $R_s^{(z)}(t) \equiv X^{(z)}(t+s) - X^{(z)}(t)$ for $s,t,z \geq 0$.

*Corollary 1: The assumptions of Proposition 6 also imply that*

$$0 \leq p\left(R_{sT}^{(z)}(tT),R_{sT}^{(z)}\big((t+u)T\big)\right)=\begin{cases}1-\dfrac{\alpha}{\beta sT}+O\left(\dfrac{1}{T^\gamma}\right), & \gamma>1 \text{ and } \beta>0 \\[3mm] 1+O\left(\dfrac{1}{T^\gamma}\right), & 0<\gamma\leq 1 \text{ and } \beta>0 \\[3mm] O\left(\dfrac{1}{T^\gamma}\right), & \gamma>0 \text{ and } \beta=0.\end{cases}\qquad(3.1.7)$$

and





$$0 \leq p\left(\left(R_{sT}^{(z)}(tT)\right)^2, \left(R_{sT}^{(z)}((t+u)T)\right)^2\right) = \begin{cases} 1 - \dfrac{2\alpha}{\beta sT} + O\left(\dfrac{1}{T^\gamma}\right), & \gamma > 1 \text{ and } \beta > 0 \\[2mm] 1 + O\left(\dfrac{1}{T^\gamma}\right), & 0 < \gamma \leq 1 \text{ and } \beta > 0 \\[2mm] O\left(\dfrac{1}{T^\gamma}\right), & \gamma > 0 \text{ and } \beta = 0. \end{cases} \qquad (3.1.8)$$

*for any $z, s, t > 1$, $u \geq s$, and sufficiently large $T$.*

Consistently with the definitions in Cont [35], the quantities in (3.1.7) are the autocorrelation coefficients for time scale $sT$ as defined for log returns with reference $z$. The quantities in (3.1.8) are the autocorrelation coefficients for time scale $sT$ as defined for squared log returns with reference $z$. Corollary 1 states that both types of autocorrelation coefficients are non-negative under the model assumptions and strictly positive for sufficiently large $T$ when $\beta > 0$. The conclusions in (3.1.8) for the cases in which $\beta > 0$ are consistent with observations of volatility clustering. The conclusions in (3.1.7) that autocorrelations of the log-returns themselves must be non-negative is more surprising, but is consistent with observations of stock indices as discussed in Section 1.2. In Section 3.3, we discuss a generalization covering the case of negative autocorrelations.

Propositions 6 and Corollary 1 describe unconditional log returns as they are commonly defined in empirical studies including Cont [35]. The present value of a derivative at a given epoch in time is determined not by the distribution of log returns defined in that way, but by the conditional distribution of log returns given the price history of the underlying up to that time epoch. The final proposition of this section describes this conditional distribution. To state that result, let $\Rightarrow$ denote convergence in distribution, and let $N(\boldsymbol{\psi}, \boldsymbol{\theta})$ denote a multivariate normal random variable for which $\boldsymbol{\psi}$ is the mean vector with $i^{th}$ component $\psi_i$ and $\boldsymbol{\theta}$ is the covariance matrix with $(i,j)^{th}$ component $\theta_{i,j}$.





*Proposition 7: Under the conditions of Proposition 6, if the parameter $\beta$ in the conclusion of Proposition 6*

*is strictly positive, and if, for any $z > 0$ and $0 < t_1 \leq t_2 \leq \cdots \leq t_N$,*

$$\frac{\int_{zT}^{(z+t_i)T} \frac{\sigma_V(S(t),t)^2 dt}{\left(1 + \frac{\beta}{\alpha^2} E \int_0^t \sigma_V(S(w),w)^2 dw\right)^2}}{E \int_{zT}^{(z+t_i)T} \frac{\sigma_V(S(t),t)^2 dt}{\left(1 + \frac{\beta}{\alpha^2} E \int_0^t \sigma_V(S(w),w)^2 dw\right)^2}} = 1 + O\left(\frac{1}{T^\gamma}\right) \tag{3.1.9}$$

*for sufficiently large $T$ (where $\gamma > 0$ is the same parameter as in condition (iii) of Proposition 6), then*

$$\left(\frac{\bar{X}^{(0)}\left((z+t_1)T\right)}{T^{1/2}}, \frac{\bar{X}^{(0)}\left((z+t_2)T\right)}{T^{1/2}}, \ldots, \frac{\bar{X}^{(0)}\left((z+t_N)T\right)}{T^{1/2}} \,\middle|\, \frac{\bar{X}^{(0)}(zT)}{T^{1/2}} = x\right) \Rightarrow N\left(\boldsymbol{\psi}(x), \boldsymbol{\theta}(x)\right) \text{ as } T \to \infty,$$

*where*

$$\psi_i(x) = \frac{z+t_i}{z} x \text{ and } \theta_{i,j}(x) = \frac{t_i(z+t_j)}{z}\alpha. \tag{3.1.10}$$

Proposition 7 can be used to approximate the distribution or moments of the present value of portfolios.

To illustrate for the one-dimensional case, Proposition 7 suggests the approximation,

$$\left(\bar{X}^{(0)}\left((z+t)T\right) \,\middle|\, \bar{X}^{(0)}(zT)\right) \approx N\left(\frac{z+t}{z}\bar{X}^{(0)}(zT), \frac{\alpha t(z+t)}{z}T\right). \tag{3.1.11}$$

Using (3.1.3), we see that (3.1.11) is itself equivalent to the log normal approximation,

$$\left(\log S\left((z+t)T\right) \,\middle|\, S(0), S(zT)\right)$$

$$\approx N\left(\log S(0) + \lambda^{(0)}\left((z+t)T\right)\right.$$

$$\left. + \frac{z+t}{z}\left(\log \frac{S(zT)}{S(0)} - \lambda^{(0)}(zT)\right), \frac{\alpha t(z+t)}{z}T\right). \tag{3.1.12}$$

The dependence of (3.1.12) not only on $S(zT)$ but also on $S(0)$ seemingly contradicts the Markov

property of $S$ implied by SDE representation in (3.1.1). We resolve the apparent paradox in Section 3.3.





If $t$ is small relative to $z$, as would be true if the forecast horizon is small relative to the prior history on which the forecast is based, then (3.1.12) implies that

$$\left(\log\frac{S\left((z+t)\mathrm{T}\right)}{S\left(z\mathrm{T}\right)} \,\middle|\, S(0), S\left(z\mathrm{T}\right)\right) \approx N\left(\lambda^{(z\mathrm{T})}(t\mathrm{T}),\ \alpha t T\right)$$

$$\approx N\left(E\int_{zT}^{(z+t)T}\mu_V(S(w),w)dw - \frac{1}{2}\alpha tT, \alpha tT\right). \qquad (3.1.13)$$

where the final expression of (3.1.13) follows from (3.1.2) and condition (iii) of Proposition 6. Section 8.5 on pages 121-124 of Ross [32] presents a mean-variance analysis -- applicable to (3.1.13) -- for the present value of positions in European call options and the underlying security when the underlying security price has a log-normal distribution. An assumption or estimate is required for the expected return $E\int_{zT}^{(z+t)T}\mu_V(S(w),w)dw$, but only over large intervals.

The limit in Proposition 7 depends on characteristics of centered log returns only through the parameter $\alpha$ from (1.1.3). Proposition 6 suggests that care must be taken if estimating $\alpha$ from covariances of log returns centered by the unconditional mean, as $\alpha$ does not appear on the right-hand side of (3.1.6) for one case in which $\beta > 0$. When $\beta > 0$, Proposition 7 suggests that estimates of $\alpha$ may be obtained more generally from estimates of log returns centered by the conditional mean.

## 3.2   Canonical Representation and Proofs.

Our final proposition provides a canonical representation for $\bar{X}^{(z)}(\cdot)$ as defined in (3.1.3) under a subset of the assumptions from Proposition 6.

*Proposition 8: If assumptions (i) and (ii) of Proposition 6 hold, then $\bar{X}^{(z)}(\cdot)$ defined in (3.1.2) satisfies*

$$\bar{X}^{(z)}(t) = \bar{X}^{(0)}(z+t) - \bar{X}^{(0)}(z) \qquad (3.2.1)$$

for any $z \geq 0$ and $t \geq 0$ *where*





$$\bar{X}^{(0)}(t) = g(t) \int_0^t \frac{\sigma_V(S(w), w)}{g(w)} \, dB(w) \qquad (3.2.2)$$

*and*

$$g(t) \equiv 1 + (\beta/\alpha^2) E \int_0^t \sigma_V(S(w), w)^2 \, dw \qquad (3.2.3)$$

*for some non-zero constant* $\beta > -\alpha^2 / E \int_0^\infty \sigma_V(S(w), w)^2 \, dw$. *The covariance function* $r^{(z)}(\cdot, \cdot)$ *defined by*

*(3.1.4) then satisfies*

$$r^{(z)}(t_1, t_2) = \left( E \int_z^{z+t_1} \sigma_V(S(w), w)^2 \, dw \right) \left( 1 + \frac{\beta}{\alpha^2} E \int_z^{z+t_2} \sigma_V(S(w), w)^2 \, dw \right). \qquad (3.2.4)$$

*Proof:*  We easily deduce (3.2.1) from (3.1.2). Using (3.1.1)-(3.1.3) and applying Ito's formula,

$$d\bar{X}^{(0)}(t) = \varphi_V(S(t), t) dt + \sigma_V(S(t), t) dB(t) \text{ for } t \geq 0, \qquad (3.2.5)$$

where $\varphi_V(s, t) \equiv \mu_V(s, t) - E\mu_V(S(t), t) - (\sigma_V(s, t)^2 - E\sigma_V(S(t), t)^2)/2$, and $\bar{X}^{(0)}(0) = 0$. By

condition (ii) of Proposition 6, (3.1.5) holds when $z = 0$ for some function $a^{(0)}(\cdot, \cdot)$. Then, by (3.3) of

Mandrekar [28], there exists a function $f(\cdot)$ that never vanishes such that

$$a^{(0)}(s, t) = f(t)/f(s) \text{ for } 0 \leq s \leq t. \qquad (3.2.6)$$

It then follows from (3.1.5) and (3.2.6) that the process

$$Y^{(0)}(t) \equiv \bar{X}^{(0)}(t)/f(t) \text{ for } t \geq 0 \qquad (3.2.7)$$

satisfies

$$\hat{E}_s^{(0)}[Y^{(0)}(t)] = Y^{(0)}(s) \text{ for all } 0 \leq s \leq t. \qquad (3.2.8)$$

A process with this property is called a *wide-sense martingale;* see pages 91 and 164-169 and of Doob

[29] for background.





By (3.2.6) and condition (ii) of Proposition 6, $f(\cdot)$ must be continuously differentiable, and using (3.2.6) we see that we can then take $f(\cdot)$ to be strictly positive without loss of generality. By (3.2.5), (3.2.7), and Ito's formula,

$$dY^{(0)}(t) = \left(\frac{\varphi_V(S(t),t) - Y^{(0)}(t)f'(t)}{f(t)}\right)dt + \frac{\sigma_V(S(t),t)}{f(t)}dB(t) \text{ for } t \geq 0, \tag{3.2.9}$$

where $Y^{(0)}(0) = 0$.

Since

$$E\left[\big(B(t) - B(s)\big)B(r)\right] = 0 \ \text{ for all } 0 \leq r \leq s \leq t,$$

it follows from arguments on page 164 of Doob [29] that

$$\hat{E}_s^{(0)}[B(t) - B(s)] = 0 \text{ for all } 0 \leq s \leq t. \tag{3.2.10}$$

From (3.2.9) and (3.2.10), we then see that

$$\lim_{\Delta s \downarrow 0} \hat{E}_s^{(0)}\left[\frac{Y^{(0)}(s + \Delta s) - Y^{(0)}(s)}{\Delta s}\right] = \frac{\varphi_V(S(s),s) - Y^{(0)}(s)f'(s)}{f(s)} \tag{3.2.11}$$

and from (3.2.8) that the left-hand side of (3.2.11) equals zero. Since $f(\cdot)$ never vanishes, we have shown that $\varphi_V(S(s),s) - Y^{(0)}(s)f'(s) = 0$ for all $s \geq 0$, so that, by (3.2.9)

$$Y^{(0)}(t) = \int_0^t \frac{\sigma_V(S(w),w)}{f(w)}dB(w) \text{ for } t \geq 0. \tag{3.2.12}$$

Equivalently,

$$Y^{(0)}(t) = B^*\left(\left[Y^{(0)}\right]_{0,t}\right) = B^*\left(\int_0^t \frac{\sigma_V(S(w),w)^2}{f(w)^2}dw\right) \tag{3.2.13}$$

for some Brownian motion $B^*$ where the first equality of (3.2.13) follows from Theorem 34.1 on page 64 of Rogers and Williams [50] and the second equality from (3.2.12) and Lemma 4.4.4 on page 143 of





Schreve [6]. It is well known that $E[B^*(s)B^*(t)] = s$ $for$ $0 \leq s \leq t$, so that, by (3.2.13) and the law of iterated expectations,

$$E[Y^{(0)}(t_1)Y^{(0)}(t_2)] = E \int_0^{t_1} \frac{\sigma_V(S(t),t)^2}{f(t)^2} dt \qquad (3.2.14)$$

for $0 \leq t_1 \leq t_2$. From (3.1.4), (3.2.7), and (3.2.14), we conclude that

$$r^{(0)}(t_1,t_2) = G(t_1)f(t_1)f(t_2) \ for \ 0 \leq t_1 \leq t_2 \qquad (3.2.15)$$

where

$$G(t) = E \int_0^t \frac{\sigma_V(S(w),w)^2}{f(t)^2} dw. \qquad (3.2.16)$$

For given $z \geq 0$, and any $0 \leq t_1 \leq t_2$, let $y^{(z)}(t_1,t_2) \equiv r^{(z)}(t_1,t_2)/r^{(z)}(t_1,t_1)$, where $r^{(z)}(\cdot,\cdot)$ is defined as in (3.1.4). By Theorem 8.1 on page 233 of Doob [29], $X^{(z)}$ has the wide-sense Markov property as assumed in condition (ii) of Proposition 6 if and only if

$$q^{(z)}(t_1,t_2,t_3) \equiv y^{(z)}(t_1,t_2)y^{(z)}(t_2,t_3) - y^{(z)}(t_1,t_3) = 0 \ for \ 0 \leq t_1 \leq t_2 \leq t_3. \qquad (3.2.17)$$

Using (3.2.15), we indeed verify that $q^{(0)}(t_1,t_2,t_3) = 0$ $for$ $0 \leq t_1 \leq t_2 \leq t_3$. For general $z \geq 0$, we obtain

$$r^{(z)}(t_1,t_2) = E\bar{X}^{(0)}(z+t_1)\bar{X}^{(0)}(z+t_2) - E\bar{X}^{(0)}(z)\bar{X}^{(0)}(z+t_1)$$

$$-E\bar{X}^{(0)}(z)\bar{X}^{(0)}(z+t_2) + E\bar{X}^{(0)}(z)\bar{X}^{(0)}(z)$$

$$= r^{(0)}(z+t_1,z+t_2) - r^{(0)}(z,z+t_1) - r^{(0)}(z,z+t_2) + r^{(0)}(z,z). \qquad (3.2.18)$$

using (3.1.4) and (3.2.1). By (3.2.17),

$$\frac{d}{dt_2}\left(\frac{d}{dz}q^{(z)}(t_1,t_2,t_3)\Big|_{z=0}\right)\Big|_{t_2=t_1} = 0 \ for \ 0 \leq t_1 \leq t_3. \qquad (3.2.19)$$

Using (3.2.17) and (3.2.18) , we find that (3.2.19) has a non-trivial solution only when





$$f(t_1)\big(f(t_1) - f(0)\big)G'(t_1) - f(0)f'(t_1)G(t_1) = 0 \ \text{ for } \ t_1 \geq 0.$$

This differential equation has two solutions:

$$f(t_1) = f(0) \ \text{ for } \ t_1 \geq 0, \qquad\qquad (3.2.20)$$

and

$$G(t_1) = c\left(1 - \frac{f(0)}{f(t_1)}\right) \ \text{ for } \ t_1 \geq 0. \qquad\qquad (3.2.21)$$

where $c$ is a non-zero constant. From (3.2.16), we see that (3.2.21) is equivalent to

$$E\int_0^{t_1} \frac{\sigma_V(S(w),w)^2}{f(w)^2} \, dw = c\left(1 - \frac{f(0)}{f(t_1)}\right) \ \text{ for } \ t_1 > 0. \qquad\qquad (3.2.22)$$

Pulling the expectation functional inside the integral sign on the left-hand side of (3.2.22), and

differentiating both sides, we obtain

$$E\sigma_V(S(t_1),t_1)^2 = cf(0)f^{'}(t_1) \ \text{ for } \ t_1 \geq 0.$$

Solving for $f(\cdot)$ and (without loss of generality) setting $c = \alpha^2/\big(\beta f^2(0)\big)$ where $\beta$ is a non-zero

constant, we obtain

$$f(t) = f(0) + \frac{\beta f(0)}{\alpha^2} E\int_0^t \sigma(S(t_1),t_1)^2 dt_1. \qquad\qquad (3.2.23)$$

Since $f(\cdot)$ in (3.2.23) must always remain non-negative, $\beta$ must satisfy the inequality in the statement of

Proposition 8. Comparing the solutions in (3.2.20) and (3.2.23), we conclude that (3.2.23) represents the

general case if we remove the restriction that $\beta$ is a non-zero.

Substituting (3.2.16) and (3.2.23) into (3.2.15), we obtain

$$r^{(0)}(t_1,t_2) = \left(E\int_0^{t_1} \sigma_V(S(w),w)^2 \, dw\right)\left(1 + \frac{\beta}{\alpha^2}E\int_0^{t_2} \sigma_V(S(w),w)^2 \, dw\right) \qquad (3.2.24)$$





for all $0 \leq t_1 \leq t_2$, and we arrive at (3.2.4) by substituting (3.2.24) into (3.2.18). The canonical representation in (3.2.2)-(3.2.3) follows from (3.2.7), (3.2.12), and (3.2.23). ∎

Since it is well known that $O\left(\frac{1}{hT}\right)^\gamma = O\left(\frac{1}{T}\right)^\gamma$ for any $h > 0$ and all $T$ sufficiently large, condition (iii) of Proposition 8 implies that there exist $\alpha > 0$ and $\gamma > 0$ such that

$$P\left(\int_z^{z+hT} \sigma_V^2(S_V(w), w)\, dw = \alpha hT\left(1 + O\left(\frac{1}{T}\right)^\gamma\right)\right) = 1 \qquad (3.2.25)$$

for any $z \geq 0$, $h > 0$, and sufficiently large $T$. Using (3.2.4) and (3.2.25), we easily reach the conclusions of Propositions 6 and Corollary 1 through formal manipulation of the big-O estimates.

To prove Proposition 7, we note that, under its assumptions, $\beta > 0$ and the conclusions of Proposition 8 hold. Since

$$E\int_z^{z+t} \frac{\sigma_V(S(w), w)^2}{g(w)^2}\, dw = \frac{\alpha^2}{\beta}\left(\frac{1}{g(z)} - \frac{1}{g(z+t)}\right)$$

for any $z, t \geq 0$, as can be verified by differentiating both sides and applying (3.2.3), we see from (3.1.9) that

$$\int_{zT}^{(z+t)T} \frac{\sigma_V(S(w), w)^2}{g(w)^2}\, dw = \frac{\alpha^2}{\beta}\left(\frac{1}{g(zT)} - \frac{1}{g((z+t)T)}\right)\left(1 + O\left(\frac{1}{T^\gamma}\right)\right) \qquad (3.2.26)$$

Applying the same logic as in (3.2.13), we also see that (3.2.2) implies that

$$\bar{X}^{(0)}(t) = g(t)B^*\left(\int_0^t \frac{\sigma_V(S(w), w)^2}{g(w)^2}\, dw\right) \qquad (3.2.27)$$

for some Brownian motion $B^*$. We will use the well-known property that Brownian motion is a Gaussian process with joint normal density





$$P(B^*(w_1) \in dy_1, \ldots, B^*(w_N) \in dy_N)$$

$$= \frac{exp\left(-\frac{1}{2}\left(\frac{y_1^2}{w_1} + \frac{(y_2-y_1)^2}{w_2-w_1} + \cdots + \frac{(y_N-y_{N-1})^2}{w_N-w_{N-1}}\right)\right)}{(\sqrt{2\pi})^N \sqrt{w_1(w_2-w_1)\cdots(w_N-w_{N-1})}} dy_1 \ldots dy_N \qquad (3.2.28)$$

for any $N \geq 1$ and $0 < w_1 < w_2 \leq \cdots \leq w_N$. We will say that that $\boldsymbol{X} \sim N(\boldsymbol{\psi}_1 \leq \boldsymbol{\psi} \leq \boldsymbol{\psi}_2, \boldsymbol{v}_1 \leq \boldsymbol{v} \leq \boldsymbol{v}_2)$ if $\boldsymbol{X}$ is a random vector with distribution function $H(\cdot)$ satisfying

$$min_{\substack{\boldsymbol{\psi}_1 \leq \boldsymbol{\psi} \leq \boldsymbol{\psi}_2 \\ \boldsymbol{v}_1 \leq \boldsymbol{v} \leq \boldsymbol{v}_2}} F(\boldsymbol{x}; \boldsymbol{\psi}, \boldsymbol{v}) \leq H(\boldsymbol{x}) \leq max_{\substack{\boldsymbol{\psi}_1 \leq \boldsymbol{\psi} \leq \boldsymbol{\psi}_2 \\ \boldsymbol{v}_1 \leq \boldsymbol{v} \leq \boldsymbol{v}_2}} F(\boldsymbol{x}; \boldsymbol{\psi}, \boldsymbol{v})$$

for each $\boldsymbol{x}$, where $F(\cdot\,; \boldsymbol{\psi}, \boldsymbol{v})$ is the cdf of a normal random vector with mean vector $\boldsymbol{\psi}$ and covariance matrix $\boldsymbol{v}$. We will use big-oh notation to specify such parameter ranges implicitly

Suppose now that $0 < t_1 \leq t_2 \leq \cdots \leq t_N$ for some $N \geq 1$. For any $1 \leq i \leq N$,

$$P\left(\frac{\bar{X}^{(0)}\left((z+t_i)T\right)}{T^{1/2}} \leq x_i \,\Big|\, \frac{\bar{X}^{(0)}(zT)}{T^{1/2}} = x\right)$$

$$= P\left(B^*\left(\int_0^{zT}\frac{\sigma(S(w),w)^2}{g(w)^2}dw + \int_{zT}^{(z+t_i)T}\frac{\sigma(S(w),w)^2}{g(w)^2}dw\right) \leq \frac{x_iT^{1/2}}{g((z+t_i)T)} \,\Big|\, B^*\left(\int_0^{zT}\frac{\sigma(S(w),w)^2}{g(w)^2}dw\right) = \frac{xT^{1/2}}{g(zT)}\right)$$

$$= P\left(B^*\left(\int_0^{zT}\frac{\sigma(S(w),w)^2}{g(w)^2}dw\right) + B^*\left(\int_{zT}^{(z+t_i)T}\frac{\sigma(S(w),w)^2}{g(w)^2}dw\right) \leq \frac{x_iT^{1/2}}{g((z+t_i)T)} \,\Big|\, B^*\left(\int_0^{zT}\frac{\sigma(S(w),w)^2}{g(w)^2}dw\right) = \frac{xT^{1/2}}{g(zT)}\right)$$

$$= P\left(B^*\left(\int_{zT}^{(z+t_i)T}\frac{\sigma(S(w),w)^2}{g(w)^2}dw\right) \leq \frac{x_iT^{1/2}}{g((z+t_i)T)} - \frac{xT^{1/2}}{g(zT)}\right)$$

$$= \int F\left(\frac{x_iT^{\frac{1}{2}}}{g((z+t_i)T)} - \frac{xT^{\frac{1}{2}}}{g(zT)}; 0, v\right) P_V\left(\int_{zT}^{(z+t_i)T}\frac{\sigma(S(w),w)^2}{g(w)^2}dw \in dv\right) \qquad (3.2.29)$$

where the first equality follows from (3.2.27), the second from the strong Markov property of Brownian motion (cf. Theorem 1 on page 5 of Harrison [51]), the third from the definition of a conditional expectation, and the last from (3.2.28). The integral in the final expression of (3.2.29) has a finite domain that is implicitly defined by (3.2.26). Using (3.2.3), (3.2.26), and (3.2.29), we obtain

$$\left(\frac{\bar{X}^{(0)}\left((z+t_i)T\right)}{T^{1/2}} \leq x_i \,\Big|\, \frac{\bar{X}^{(0)}(zT)}{T^{1/2}} = x\right) \sim N(\psi_i, v_{ii}) \qquad (3.2.30)$$

where





$$\psi_i = \begin{cases} \left(\frac{z+t_i}{z} - \frac{\alpha t_i}{\beta z^2 T} + O\left(\frac{1}{T^\gamma}\right)\right)x, \; if \; \gamma > 1 \\ \left(\frac{z+t_i}{z} + O\left(\frac{1}{T^\gamma}\right)\right)x, \; if \; 0 < \gamma \le 1 \end{cases} \quad and \quad v_{i,i} = \begin{cases} \frac{\alpha t_i(z+t_i)}{z} - \frac{\alpha^2 t_i^2}{\beta z^2 T} + O\left(\frac{1}{T^\gamma}\right), \; if \; \gamma > 1 \\ \frac{\alpha t_i(z+t_i)}{z} + O\left(\frac{1}{T^\gamma}\right), \; if \; 0 < \gamma \le 1. \end{cases} \quad (3.2.31)$$

For any $N \ge 2$ and $0 < t_1 \le t_2 \le \cdots \le t_N$, the steps leading to the expression on the left-hand side of the last equality of (3.2.29) generalize to show that $\left(\frac{\bar{X}^{(0)}\left((z+t_1)T\right)}{T^{1/2}}, \dots, \frac{\bar{X}^{(0)}\left((z+t_N)T\right)}{T^{1/2}} \mid \frac{\bar{X}^{(0)}(zT)}{T^{1/2}} = x\right)$ is the joint distribution of Brownian motion at successive times $\int_{zT}^{(z+t_i)T} \frac{\sigma(S(w),w)^2}{g(w)^2} dw$ for $i = 1, \dots, N$. Using (3.2.26) to generalize the final expression of (3.2.29) and noting that the domain of the integral is again finite, we conclude that

$$\left(\frac{\bar{X}^{(0)}\left((z+t_1)T\right)}{T^{1/2}}, \dots, \frac{\bar{X}^{(0)}\left((z+t_N)T\right)}{T^{1/2}} \mid \frac{\bar{X}^{(0)}(zT)}{T^{1/2}} = x\right) \sim N(\boldsymbol{\psi}_1 \le \boldsymbol{\psi} \le \boldsymbol{\psi}_2, \boldsymbol{v}_1 \le \boldsymbol{v} \le \boldsymbol{v}_2) \qquad (3.2.32)$$

for some $\boldsymbol{\psi}_1 \le \boldsymbol{\psi} \le \boldsymbol{\psi}_2$ and $\boldsymbol{v}_1 \le \boldsymbol{v} \le \boldsymbol{v}_2$. The bounds on $i^{th}$ component of $\boldsymbol{\psi}$ must agree with the implicit bounds for the mean value in (3.2.31) as was derived for the one-dimensional case, and the bounds on the $(i,j)^{th}$ component of $\boldsymbol{v}$ must agree with

$$v_{i,j} = \begin{cases} \frac{\alpha t_i(z+t_j)}{z} - \frac{\alpha^2 t_i t_j}{\beta z^2 T} + O\left(\frac{1}{T^\gamma}\right), & if \; \gamma > 1 \\ \frac{\alpha t_i(z+t_j)}{z} + O\left(\frac{1}{T^\gamma}\right), & if \; 0 < \gamma \le 1 \end{cases} \qquad (3.2.33)$$

as is easily derived for the two-dimensional case. We verify that (3.2.33) agrees with (3.2.31) when $j = i$. Proposition 7 follows from (3.2.31)-(3.2.33) by taking limits as $T \to \infty$.

Proposition 7 is an expression of aggregational normality. By similar arguments, aggregational normality also holds for unconditional log returns.

## 3.3  A Closer Look at Autocorrelations

By (3.2.2),

$$d\bar{X}^{(0)}(t) = \frac{g'(t)}{g(t)}\bar{X}^{(0)}(t)dt + \sigma_V(S(t),t)dB(t) \; for \; t \ge 0 \qquad (3.3.1)$$





where $\bar{X}^{(0)}(0) = 0$. Since $g(\cdot)$ is positive and $g'(\cdot)$ is non-negative under the assumptions of Proposition 6, the first summand on the right-hand side of (3.3.1) is responsible for the the non-negative autocorrelation structure for log returns seen in (3.1.6) and (3.1.7) of Corollary 2. As the parameter $\beta$ in (3.2.3) approaches zero, so does $g'(\cdot)$; and autocorrelations approach zero.

When (3.3.1) holds,

$$\mu(s,t) = \frac{g'(t)}{g(t)}\left(\log\left(\frac{s}{S(0)}\right) - \lambda_0(t)\right) - \mathrm{d}\lambda^{(0)}(t)/dt + \frac{\sigma^2(s,t)}{2} \, for \, s > 0 \, and \, t \geq 0. \qquad (3.3.2)$$

in (3.1.1). Consequently, the conditional distribution of $\{S(t): t \geq w\}$ given $S(w)$ will depend on $S(0)$ through $\mu(\cdot,\cdot)$ even though $S$ is a Markov process. The dependence of (3.3.2) on $S(0)$ vanishes along with the positive autocorrelations of $\bar{X}^{(0)}$ as $\beta$ approaches zero.

The conclusion in Proposition 6 that $\beta$ must be non-negative depends on our assumptions that the time domain for the price process $S$ is the full half line and that $E \int_0^\infty \sigma_V(S(t),t)^2 dt = \infty$. In the more general setting of Proposition 8, $\beta$ can take on negative values, but only if $E \int_0^\infty \sigma_V(S(t),t)^2 dt < \infty$. Cases in which $\beta < 0$ can be modeled for more realistic scenarios for which $E \int_0^\infty \sigma_V(S(t),t)^2 dt = \infty$ by constraining the time domain for $S$ to a bounded interval. If the time domain for $S$ is restricted to an interval $[0, R]$ where $R$ is sufficiently large, then the conclusions of Proposition 8 will continue to hold on $[0, R]$ if $\beta > -\alpha^2/E \int_0^R \sigma_V(S(t),t)^2 dt$. The asymptotic estimates obtained in Propositions 6 and Corollary 1 also are easily generalized to hold on that finite interval. Corollary 1 generalizes to show that autocorrelations of log returns can exhibit negative autocorrelations over a finite interval, but that autocorrelations of squared log returns are always non-negative over both finite and infinite intervals.